\documentclass[twocolumn,prl,showpacs,superscriptaddress,amsmath,amssymb]{revtex4-1}

\usepackage{graphicx}
\usepackage{dcolumn}
\usepackage{bm}

\begin{document}

\title{Electron-phonon coupling in the conventional superconductor YNi$_2$B$_2$C at high phonon energies studied by time-of-flight neutron spectroscopy}

\author{F. Weber}
\email{frank.weber@kit.edu}
\affiliation{Materials Science Division, Argonne National Laboratory, Argonne, Illinois 60439, USA}
\affiliation{Karlsruhe Institute of Technology, Institute of Solid State Physics, P.O. Box 3640, D-76021 Karlsruhe, Germany}
\author{S. Rosenkranz}
\affiliation{Materials Science Division, Argonne National Laboratory, Argonne, Illinois 60439, USA}
\author{L. Pintschovius}
\affiliation{Karlsruhe Institute of Technology, Institute of Solid State Physics, P.O. Box 3640, D-76021 Karlsruhe, Germany}
\author{J.-P. Castellan}
\author{R. Osborn}
\affiliation{Materials Science Division, Argonne National Laboratory, Argonne, Illinois 60439, USA}
\author{W. Reichardt}
\author{R. Heid}
\author{K.-P. Bohnen}
\affiliation{Karlsruhe Institute of Technology, Institute of Solid State Physics, P.O. Box 3640, D-76021 Karlsruhe, Germany}
\author{E. A. Goremychkin}
\affiliation{Materials Science Division, Argonne National Laboratory, Argonne, Illinois 60439, USA}
\author{A. Kreyssig}
\altaffiliation[present address: ]{Ames Laboratory and Department of Physics and Astronomy, Iowa State University, Ames, IA, 50011, USA}
\affiliation{Technische Universit\"at Dresden, Institut f\"ur Festk\"orperphysik, D-01062 Dresden, Germany}
\author{K. Hradil}
\altaffiliation[present address: ]{Technische Universit\"at Wien, R\"ontgenzentrum, A- 1060 Wien, Austria}
\affiliation{Universit\"at G\"ottingen, Institut f\"ur Physikalische Chemie, Au{\ss}enstelle FRM-II, D-85747 Garching, Germany}
\author{D. L. Abernathy}
\affiliation{Neutron Scattering Sciences Division, Oak Ridge National Laboratory, Oak Ridge, Tennessee 37831, USA}

\date{\today}

\begin{abstract}
  We report an inelastic neutron scattering investigation of phonons with energies up to $159\,\rm{meV}$ in the conventional superconductor YNi$_2$B$_2$C. Using the SWEEP mode, a newly developed time-of-flight technique involving the continuous rotation of a single crystal specimen, allowed us to measure a four dimensional volume in ($\mathbf{Q},E$) space and, thus, determine the dispersion surface and linewidths of the $A_{1g}$ ($\approx 102\,\rm{meV}$) and $A_u$ ($\approx 159\,\rm{meV}$) type phonon modes for the whole Brillouin zone. Despite of having linewidths of $\Gamma = 10\,\rm{meV}$, $A_{1g}$ modes do not strongly contribute to the total electron-phonon coupling constant $\lambda$. However, experimental linewidths show a remarkable agreement with \textit{ab-initio} calculations over the complete phonon energy range demonstrating the accuracy of such calculations in a rare comparison to a comprehensive experimental data set.
\end{abstract}



\maketitle

Superconductivity is one of the most intriguing phenomena in condensed matter physics and each discovery of new compounds, such as the iron based superconductors, sparks a tremendous amount of scientific interest.  To date, BCS theory \cite{Bardeen57}, with the Cooper pairing mechanism provided by electron-phonon interaction, represents the only well-understood and experimentally verified microscopic picture of superconductivity, and systems which are successfully described by BCS theory are known as conventional superconductors. The corresponding superconducting transition temperature can be calculated by, e.g., the Allen-Dynes formula \cite{Allen75}:
\begin{equation}\label{equ1}
T_c = \frac{\omega_{eff}}{1.2}exp\left(\frac{-1.04(1+\lambda)}{\lambda-\mu^{*}(1+0.62\lambda)}\right)
\end{equation}
Here, $\lambda$ is the electron-phonon coupling (EPC) constant, $\mu^{*}$ the effective electron-electron interaction and $\omega_{eff}$ the effective phonon frequency. Thus, a large effective phonon frequency and electron-phonon coupling constant are favorable for a high superconducting transition temperature. Determining $\lambda$ and $\omega_{eff}$ microscopically, i.e. by measuring all phonon energies and linewidths, is extremely time consuming if not impossible. Therefore, \textit{ab-initio} calculations of the lattice dynamical properties are often used to estimate $\lambda$ and the corresponding superconducting transition temperature $T_c$\cite{Boeri08}.

Following the discovery of superconductivity in the rare-earth nickel-borocarbides (RNBC) \cite{Cava94}, the conventional phonon mediated pairing mechanism was quickly accepted to be valid for these compounds, based on density-functional-theory (DFT) predicting a large $\lambda = 2.6$ \cite{Pickett94} and the observation of a strong isotope effect \cite{Lawrie95,Cheon99}. The calculated electronic structure \cite{Pickett94} showed that Ni and B states dominate at the Fermi energy. When a strong sensitivity of $T_c$ towards the bonding angle of the NiB$_4$ tetrahedron was observed \cite{Siegrist94}, it was even suggested that high energy phonons of $A_{1g}$ symmetry, which modulate this bonding angle, were prominently involved in mediating superconductivity \cite{Pickett94,Matheiss94}. This is somewhat surprising, as, according to equation \ref{equ1}, an EPC constant of $\lambda = 1$ would only require $\omega_{eff} = 20\,\rm{meV}$ in order to explain a $T_c$ of about $15\,\rm{K}$ ($\mu^{*} = 0.13$), as it is observed in the RNBC compound YNi$_2$B$_2$C. In contrast, the frequency of the $A_{1g}$ mode is close to $100\,\rm{meV}$ \cite{Gompf97}.

\begin{figure}
\begin{center}
\includegraphics[width=0.95\linewidth]{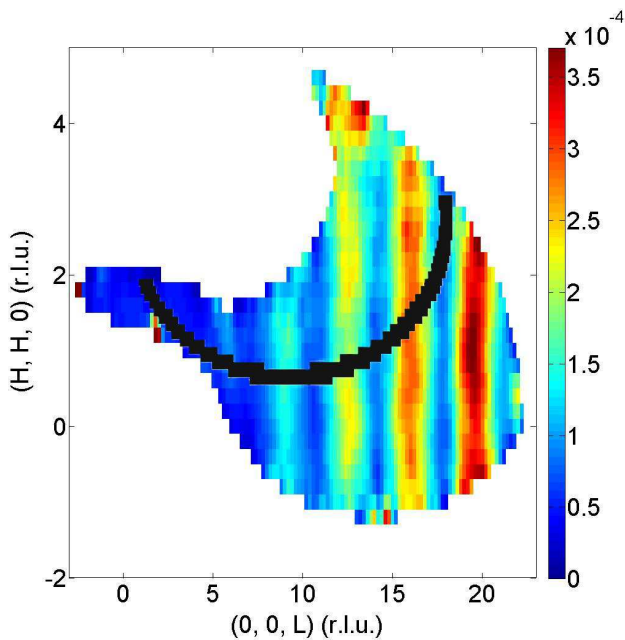}
\caption{\label{fig_1} (Color online) Raw data collected in the $(hh0)-(00l)$ plane with an incident energy of $E_i = 160\,\rm{meV}$. The figure shows a cut at $E = (102.5 \pm 2.5)\,\rm{meV}$. The data were obtained with an angular rotation of the specimen of $60^{\circ}$. The black pixels illustrate the wave vector range covered with an angular range of $1^{\circ}$ (no intensity scale applied). Note that the wave vectors covered in an experiment with such a small angular range changes significantly with energy transfer.}
\end{center}
\end{figure}

\begin{figure}
\begin{center}
\includegraphics[width=0.95\linewidth]{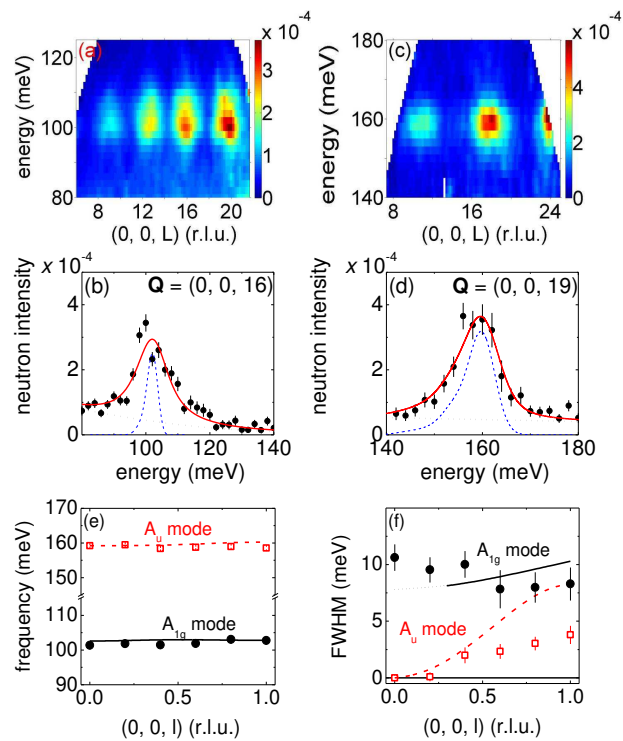}
\caption{\label{fig_2} (Color online) \textit{(a)(c)} Color-coded contour plot of raw data for varying energy transfers vs wave vectors $\mathbf{Q} = (0, 0, l)$ obtained with incident energies of $E_i = 160\,\rm{meV}$ \textit{(left)} and $250\,\rm{meV}$ \textit{(right)}.  \textit{(b)(d)} Examples of data analysis of energy scans extracted from the measured $(\mathbf{Q},E)$ space at the given wave vectors and incident energies of $E_i = 160\,\rm{meV}$ \textit{(left)} and $250\,\rm{meV}$ \textit{(right)}. Constant-\textbf{Q} scans are binned over $\pm 0.1\,\rm{r.l.u.}$ along $(00L)$ and $\pm 0.2\,\rm{r.l.u.}$ along $(HH0)$ and $(H\bar{H}0)$ and are analyzed by fitting a Lorentzian convoluted with the instrumental resolution function. Solid/dashed/dotted lines are the respective convoluted fit/resolution function/experimental background. Experimentally observed phonon frequencies and linewidths (FWHM) of the A$_{1g}$ (filled symbols) and A$_u$ (open symbols) phonon modes are shown in panels \textit{(e)} and \textit{(f)}. Solid and dashed lines were calculated from DFPT. The dotted part of the line in \textit{(f)} indicates that these values could be calculated only approximately (see text). }
\end{center}
\end{figure}

In this letter, we report measurements of the $A_{1g}$ and $A_u$ type phonons at very high energies, i.e. $E \approx 100\,\rm{meV}$ and $160\,\rm{meV}$, respectively, and another pure B-C mode (see supplemental information) dispersing at $37\,\rm{meV} \le E \le 52\,\rm{meV}$ in YNi$_2$B$_2$C. Utilizing a novel technique whereby the sample is continuously rotated, the SWEEP mode developed at the TOF chopper spectrometer ARCS \cite{Abernathy08} located at the Spallation Neutron Source (SNS), Oak Ridge National Laboratory, we were able to sample a four dimensional volume in $(\mathbf{Q},E)$ space, from which we extracted constant energy planes in reciprocal space (Fig.~\ref{fig_1}) or data for varying energy transfers and constant $\mathbf{Q}$ along all high- and off-symmetry directions (Fig.~\ref{fig_2}). Results from TOF spectroscopy are complemented by triple-axis measurements for lower energy phonon linewidths requiring an improved energy resolution. We compare our results to detailed calculations of the lattice dynamical properties based on density-functional-perturbation theory (DFPT), which recently challenged the importance of the high energy modes for superconductivity in boron carbides \cite{Reichardt05}.

Neutron scattering experiments for such high energy excitations are challenging as one has to measure a significant part of the Brillouin zone with an energy resolution comparable to the electronic contribution to the phonon linewidth. The low neutron flux at energy transfers of about $100\,\rm{meV}$ and concomitant poor energy resolution exclude the usage of thermal neutron scattering on a triple-axis spectrometer (TAS), which had nearly a monopoly on energy and momentum resolved phonon investigations up to very recent times. In the last decade, high-resolution inelastic x-ray scattering (IXS) has become a serious competitor to TAS measurements. However, phonons at high energy transfers are generally associated with displacements of the light elements, which have only small x-ray scattering amplitudes.

Time-of-flight (TOF) chopper spectrometers at modern pulsed neutron sources \cite{Watanabe03} offer a way out of this predicament. They combine a high neutron flux over a large energy range with good energy resolution at large energy transfers \cite{Abernathy08}. Furthermore, these instruments map large areas of reciprocal space and energy simultaneously employing arrays of position sensitive $^{3}$He detector tubes of up to $30\,\rm{m^2}$ \cite{Abernathy08,Bewley09}. However, the layout of such a TOF instrument with fixed detector positions covering a cylindrical surface around the sample position does not allow one to make energy scans at constant momentum transfer $\mathbf{Q}$, the usual choice in neutron TAS measurements. Instead, the instrument samples a curved surface of reciprocal space as indicated in Fig.~\ref{fig_1}, which makes it impossible to directly extract a lifetime of an observed excitation. To do so requires fitting of a full four-dimensional resolution convolution with an a-priori assumed model for the dispersion relation \cite{PerringFIT}. In addition, the range of measured wave vectors strongly depends on the energy transfer. In measurements of quasi two-dimensional, e.g., magnetic excitations, this drawback is not too serious because all results may be projected onto a basal plane. For phonon measurements however, and generally for measuring any excitation that is strongly dispersing in all three dimensions, such a projection is rarely if ever meaningful. This has stimulated the development of a new technique for the determination of phonon dispersion curves using a TOF instrument, which we call the SWEEP mode. The SWEEP mode involves a continuous rotation of the single crystal sample during the measurements so that large volumes of reciprocal space are sampled. We note that the SWEEP mode provides greater flexibility than the related HORACE mode currently being developed at ISIS \cite{PerringHOR}, where the sample is measured at a series of discrete angular positions over a period of a day or more. In the SWEEP mode, the entire volume is measured within minutes, allowing the experimenter to optimize the angular range as the data are accumulated.

The sample we used was the same single crystal as used previously in the TAS measurements of the low energy phonons \cite{Weber08,Pintschovius08} with a volume of about $0.4\,\rm{cm^3}$. Thus, all the progress made in the TOF measurements came from the technique, and not an increase in sample mass.  Data were collected using incident energies of $E_i = 160\,\rm{meV}$ and $250\,\rm{meV}$ for measuring the branches involving the $A_{1g}$ ($\approx100\,\rm{meV}$) and the $A_u$ ($\approx160\,\rm{meV}$) modes, respectively. Because these modes are flat in momentum space, their lifetimes can be easily extracted from energy cuts at constant $\mathbf{Q}$, obtained from the reconstruction of the full $S(\mathbf{Q},E)$, and only require a one-dimensional resolution convolution. The resolution function, an Ikeda-Carpenter function describing the moderator \cite{Ikeda85} convoluted with a Gaussian describing the chopper contribution, was determined at zero energy transfer from vanadium scans and scaled to finite energy transfers \cite{Windsor81}. The so determined instrumental full widths at half maximum are $4.0\,\rm{meV}$ ($E_i = 160\,\rm{meV}, \Delta E = 102\,\rm{meV}$) and $7.4\,\rm{meV}$ ($E_i = 250\,\rm{meV}, \Delta E = 159\,\rm{meV}$) (Figs.~\ref{fig_2}b, d). The sample was mounted in a closed-cycle refrigerator and measurements were done at $T = 5\,\rm{K}$. The total data acquisition time for the results presented here were 6 days. The sample was continuously rotated back and forth throughout the experiment with an angular speed of about $5^{\circ}$ per minute over a range of $60^{\circ}$ ($E_i = 160\,\rm{meV}$) or $38^{\circ}$ ($E_i = 250\,\rm{meV}$). Each detected neutron was stored with the respective sample rotation angle in order to allow a later transformation into $(\mathbf{Q},E)$ space. Linewidths of the mode dispersing at $37\,\rm{meV} \le E \le 52\,\rm{meV}$ were obtained from standard triple-axis experiments on the 1T and PUMA spectrometers located at the ORPHEE reactor at LLB, Saclay, and at the research reactor FRM II in Munich, respectively (for experimental details see supplemental information).

\begin{figure}
\includegraphics[width=0.95\linewidth]{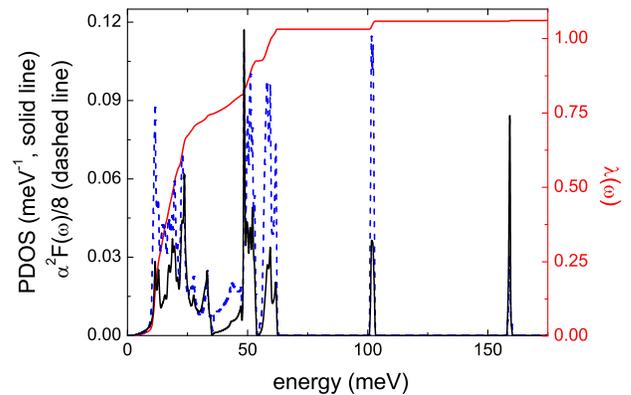}
\caption{\label{fig_3} (Color online) \textit{Left-hand scale: Ab-initio} calculations of the phonon density-of-states (black, solid line), the Eliashberg function $\alpha^2F(\omega)$ (blue, dashed line). \textit{Right-hand scale:} Partially energy integrated electron-phonon coupling constant $\lambda=2\cdot\int^{\omega}_{0}d\omega^{'}\frac{\alpha^2F(\omega^{'})}{\omega^{'}}$ (red, solid line).}
\end{figure}

Raw data obtained in TOF spectroscopy are shown in Figs.~\ref{fig_1} and \ref{fig_2}. Obviously, there is a strong variation of the scattered intensities for both of the high energy phonon branches as a function of wave vector along the line $(00L)$, whereas there is nearly no variation along $(HH0)$ (Figs.~\ref{fig_1} and \ref{fig_2}a, c). We note that the observed intensity variations quantitatively agree with that predicted by DFPT based on the calculated phonon patterns (see supplemental information). For a detailed comparison of the observed and calculated frequencies and linewidths, constant $\mathbf{Q}$ scans, binned over $\pm 0.1\,\rm{r.l.u.}$ along $(00L)$ and $\pm 0.2\,\rm{r.l.u.}$ along $(HH0)$ and $(H\bar{H}0)$, were fitted by a Lorentzian convoluted with the instrumental resolution function [wave vectors are given in reciprocal lattice units (r.l.u.) of $(2\pi/a, 2\pi/b, 2\pi/c)$, where $a = b = 3.51\,$\AA, and $c = 10.53\,$\AA]. Two examples of fits to the raw data are shown in Figs.~\ref{fig_2}(b) and (d). In this way, we investigated large sections of the Brillouin zone including the main high symmetry directions along the crystallographic $(100)$, $(110)$ and $(001)$ axes. Representatively, we show results for the $(001)$ direction in Figs.~\ref{fig_2}(e) and (f). For both phonon branches, we see good agreement between calculated and observed frequencies with deviations of less than $2\%$ over the whole Brillouin zone. What is more, the calculated and observed phonon linewidths, which are all in the range of $8 - 10\,\rm{meV}$, agree quite well for the branch starting from the $A_{1g}$ mode. Hence, there is strong EPC throughout the Brillouin zone. Why it nonetheless does not significantly contribute to $\lambda$ will be discussed below. As to the branch starting from the $A_u$-mode, theory and experiment agree well except for the $(001)$ direction: the phonon is resolution limited in most parts of the Brillouin zone, but acquires a detectable linewidth near the zone boundary along the $(001)$ direction. This is qualitatively predicted by theory, but quantitatively DFPT overestimates the phonon linewidths, i.e. DFPT will predict a larger impact of the $A_u$ modes EPC than experimentally corroborated. We now use the DFPT results for a discussion about the contribution of the various phonon modes to the total electron-phonon coupling constant $\lambda$.

\begin{figure}
  \includegraphics[width=0.95\linewidth]{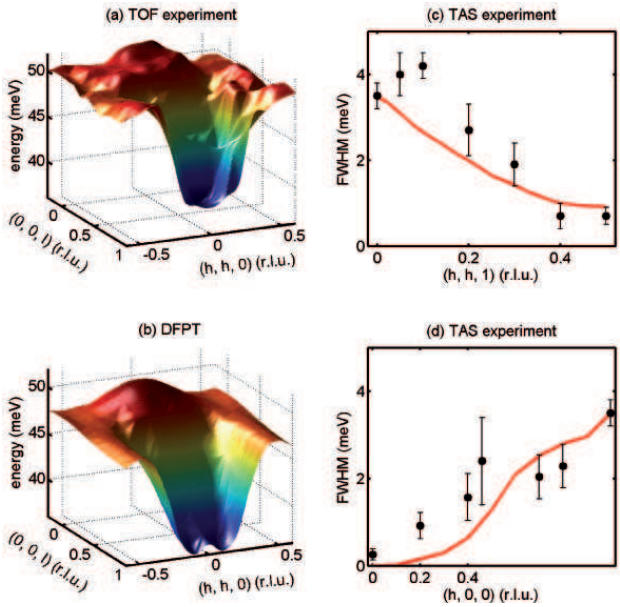}
  \caption{\label{fig_5} (Color online) Observed \textit{(a)} and calculated \textit{(b)} dispersion surface of a phonon mode showing a pronounced anomaly in the \textit{(hh0)}-\textit{(00l)} plane. Experimental phonon energies were determined from TOF data in steps of 0.1 r.l.u. in \textit{h} and \textit{l}. \textit{(c)(d)} Measured and calculated phonon linewidths of the mode shown in \textit{(a)} and \textit{(b)}, respectively, along two high symmetry lines of the Brillouin zone. Experimental results shown in \textit{(c)} and \textit{(d)} were obtained by triple-axis spectroscopy (see text).}
\end{figure}

In spite of the large phonon linewidth of the $A_{1g}$ and, to some extend, $A_u$ modes, DFPT claims that two thirds of the total $\lambda$ is due to phonons with $E \le 30\,\rm{meV}$ (Fig.~\ref{fig_3}) and in fact more than $97\%$ is due to phonons with $E \le 70\,\rm{meV}$. In order to understand why the sizeable linewidth of the $A_{1g}$ mode does not contribute more strongly to $\lambda$ and thereby to superconductivity, it is instructive to look into the different components of the Allen-Dynes formula (equation \ref{equ1}). From the definitions of the EPC coupling constant $\lambda$,

\begin{equation}\label{equ2}
\lambda=2\cdot\int^{\infty}_{0}d\omega\frac{\alpha^2F(\omega)}{\omega},
\end{equation}

and the effective phonon frequency $\omega_{eff}$,

\begin{equation}\label{equ3}
\omega_{eff}=exp\left(\frac{2}{\lambda}\cdot\int^{\infty}_{0}d\omega\frac{\alpha^2F(\omega)}{\omega}ln(\omega)\right),
\end{equation}

we see that both depend on the Eliashberg function $\alpha^2F(\omega)$ divided by the frequency. The importance of the weighting factor $1/\omega$ can be seen in Fig.~\ref{fig_3}, which shows the calculated phonon density-of-states (PDOS), $\alpha^2F(\omega)$ and also the partially energy integrated electron-phonon coupling constant $\lambda$. The $A_{1g}$ modes around $100\,\rm{meV}$ clearly have an enlarged contribution to $\alpha^2F(\omega)$ with respect to their spectral weight in the PDOS, indicative of substantial EPC. But in the end, their integrated contribution to $\lambda$ remains fairly small, much smaller than that of phonon modes below $25\,\rm{meV}$: therefore, phonons above $70\,\rm{meV}$ account only for $3\%$ of the total $\lambda$. Likewise, the value of $\omega_{eff}$ is only slightly enhanced by the high-frequency modes: the value computed from the DFPT results is $\omega_{eff}=23.16\,\rm{meV}$. Therefore, these modes are not decisive for the relatively high $T_{c,exp} = 15.6\,\rm{K}$. These conclusions are supported by the above mentioned fact that the observed linewidths of the $A_u$-related modes are even smaller than predicted by DFPT.

While DFPT clearly shows the negligible role of the high energy phonon branches and correctly predicts strong EPC in the acoustic phonon branches \cite{Weber08,Pintschovius08,Kawano96}, it also predicts an extremely steep phonon dispersion of a pure Ni-B mode dispersing from $52\,\rm{meV}$ at the zone center to $36\,\rm{meV}$ at the \textit{Z} point, i.e. $\mathbf{q} = (0, 0, 1)$. This anomalous dispersion, with a minimum centered at the \textit{Z} point, is clearly related to strong EPC (see Fig.~\ref{fig_5} and Ref.~\onlinecite{Reichardt05}). Experimentally, we were able to extract corresponding frequencies over a large section of the Brillouin zone from our data set with $E_i = 160\,\rm{meV}$. We compare the obtained dispersion surface in the $(hh0)$-$(00l)$ scattering plane, consisting of 169 individually determined frequencies, to the calculated dispersion in Fig.~\ref{fig_5}. We note that it would be very time consuming and impractical to use a conventional TAS for the same purpose even at these lower energy transfers. As can be seen from a comparison of Figs.~\ref{fig_5}(a) and (b), the experiment and theory agree quite well. Due to the limited amount of beam time we were not able to investigate the corresponding phonon linewidths requiring better energy resolution, i.e. a lower incident energy. Therefore, we show phonon linewidths along $(h, 0, 0)$ and $(h,h,1)$ for the mode dispersing between $37$ and $52\,\rm{meV}$ obtained in a TAS measurement in Fig.~\ref{fig_5}\textit{(c)} and \textit{(d)}\footnote{A detailed report of this investigation will be published elsewhere. Two exemplary  data sets are shown in the supplementary information.}. We observe good agreement between the predicted contribution to the phonon linewidth and the experimental results. Taking into account predicted and reported strong EPC in acoustic branches of YNi$_2$B$_2$C \cite{Pintschovius08}, our new results for phonon modes with $37\,\rm{meV} \le E \le 160\,\rm{meV}$ demonstrate that DFPT yields reliable predictions of the lattice dynamical properties over the whole energy and wave vector spectrum in the conventional superconductor YNi$_2$B$_2$C.

In conclusion, the recently developed SWEEP mode involving a continuous rotation of a single crystalline sample on a TOF chopper spectrometer was used to obtain very detailed information on phonon frequencies as well as phonon linewidths over a large region in reciprocal space and up to very high phonon energies using a sample of moderate volume. Using this technique, we show that the high energy vibrations in the superconductor YNi$_2$B$_2$C couple indeed very strongly to the electrons, but that this coupling is nevertheless irrelevant to superconductivity, in contrast to earlier claims \cite{Pickett94,Matheiss94} and the common association that high energy phonons with strong electron-phonon coupling yield a large effective phonon frequency $\omega_{eff}$ and, consequently, $T_c$. The latter only applies for superconductors like MgB$_2$, where nearly all of the EPC is in one high energy branch \cite{Heid02}, but it does not hold true if there are several phonons, both at low and high energies, with strong EPC, as is the case for YNi$_2$B$_2$C. Here, as presented in detail above, the effective phonon frequency is dominated by the low energy phonons, and the high energy phonons provide only a small contribution. Our results show good agreement with DFPT calculations for the phonon energies and linewidths and corroborate the DFPT result that lower energy phonon modes with large amplitudes of the lighter atoms mainly mediate superconductivity in YNi$_2$B$_2$C.

\begin{acknowledgments}
Work at Argonne was supported by U.S. Department of Energy, Office of Science, Office of Basic Energy Sciences, under contract No. DE-AC02-06CH11357. The research at Oak Ridge National Laboratory's Spallation Neutron Source was sponsored by the Scientific User Facilities Division, Office of Basic Energy Sciences, U.S. Department of Energy.
\end{acknowledgments}


\end{document}